%% file: main.tex
\PassOptionsToPackage{unicode,pdfusetitle,bookmarksnumbered}{hyperref}
\RequirePackage{hyperref}
\documentclass[acmsmall]{acmart}

\usepackage{etex}

\usepackage{subcaption} 

\let\terms\undefined

\usepackage{booktabs}   
\usepackage{subcaption} 
\usepackage[noend]{algpseudocode}

\usepackage{adjustbox}

\usepackage{standalone} 

\usepackage{pgf-pie}

\usepackage{comment}

\usepackage{threeparttable}

\input{preamble}
\author{Jialu Zhang}
\authornote{Corresponding author.}
\authornotemark[2]
\affiliation{%
  \institution{University of Waterloo}
  \city{Waterloo}
  \country{Canada}}
\email{jialu.zhang@uwaterloo.ca}

\author{Jialiang Gu}
\authornote{Equal contribution.}
\affiliation{%
  \institution{George Mason University}
  \city{Fairfax}
  \country{USA}
}
\email{jgu7@gmu.edu}

\author{Wangmeiyu Zhang}
\affiliation{%
  \institution{Chinese University of Hong Kong, Shenzhen}
  \city{Shenzhen}
  \country{China}
}
\email{123090825@link.cuhk.edu.cn}

\author{José Pablo Cambronero}
\affiliation{%
  \institution{Google}
  \city{Atlanta}
  \country{USA}
}
\email{josepablocam@gmail.com}

\author{John Kolesar}
\affiliation{%
  \institution{Yale University}
  \city{New Haven}
  \country{USA}
}
\email{john.kolesar@yale.edu}

\author{Ruzica Piskac}
\affiliation{%
  \institution{Yale University}
  \city{New Haven}
  \country{USA}
}
\email{ruzica.piskac@yale.edu}

\author{Daming Li}
\affiliation{%
  \institution{Independent Researcher}
  \city{Mountain View}
  \country{USA}
}
\email{damingliyale22@gmail.com}

\author{Hanyuan Shi}
\affiliation{%
  \institution{Independent Researcher}
  \city{Hangzhou}
  \country{China}
}
\email{shihanyuan1995@gmail.com}

\begin{document}

\title{A Systematic Study of Time Limit Exceeded Errors in Online Programming Assignments}

\begin{abstract}
	\input{sec/abs}

\end{abstract}

\maketitle

\input{sec/intro}
\input{sec/motiv}
\input{sec/survey}
\input{sec/systemdesign}
\input{sec/evaluation}
\input{sec/discussion}

\input{sec/related}
\input{sec/conclusions}

\bibliography{repair,competitions}
\end{document}

%% file: preamble.tex
\usepackage{graphicx}
\usepackage{amsmath}

\usepackage{xspace}
\usepackage{footnote}
\usepackage{wrapfig}  

\usepackage{amsfonts}
\usepackage{natbib}
\usepackage{hhline}
\usepackage{multirow}
\usepackage{setspace} 
\usepackage{url}
\usepackage{booktabs}
\usepackage{xcolor}
\usepackage{lipsum}
\usepackage{courier}
\usepackage{listings}
\usepackage{caption}
\usepackage{colortbl}
\usepackage{arydshln} 
\usepackage{makecell} 
\usepackage{tikz}
\usepackage{xcolor}
\usetikzlibrary{calc}
\usepackage{varwidth}
\usepackage{natbib}
\usepackage{algorithm}
\usepackage[noend]{algpseudocode}
\usepackage{subcaption}
\usepackage{wrapfig}

\usepackage[most]{tcolorbox}

\usetikzlibrary{fit,arrows,calc,positioning,quotes}
\usetikzlibrary{decorations.pathreplacing}

\usepackage{mathtools}
\usepackage{fancyvrb}
\usepackage{textcomp}
\usepackage{stmaryrd}

\tikzset{
  -|-/.style={
    to path={
      (\tikztostart) -| ($(\tikztostart)!#1!(\tikztotarget)$) |- (\tikztotarget)
      \tikztonodes
    }
  },
  -|-/.default=0.5,
  |-|/.style={
    to path={
      (\tikztostart) |- ($(\tikztostart)!#1!(\tikztotarget)$) -| (\tikztotarget)
      \tikztonodes
    }
  },
  |-|/.default=0.5
}

\usetikzlibrary{shapes.geometric}

\long\def\com#1{}

\newcommand{\app}{\texttt{Nettle}\xspace}
\newcommand{\appeval}{\texttt{Nettle-Eval}\xspace}

\newcommand{\overallaccuracy}{\ensuremath{98.5\%}\xspace}

\newcommand{\para}[1]{\smallskip\noindent {\bf #1}}

\newcommand{\squishlist}{
   \begin{list}{$\bullet$}
    { \setlength{\itemsep}{0pt}      \setlength{\parsep}{3pt}
      \setlength{\topsep}{3pt}       \setlength{\partopsep}{0pt}
      \setlength{\leftmargin}{3.5mm} \setlength{\labelwidth}{1em}
      \setlength{\labelsep}{0.5em} } 
}

\newcommand{\squishend}{
    \end{list}  }

\def\BibTeX{{\rm B\kern-.05em{\sc i\kern-.025em b}\kern-.08em
    T\kern-.1667em\lower.7ex\hbox{E}\kern-.125emX}}

%% file: sec/abs.tex
Online programming platforms such as Codeforces and LeetCode attract millions of users seeking to learn to program or refine their skills for industry interviews. A major challenge for these users is the Time Limit Exceeded (TLE) error, triggered when a program exceeds the execution time bound. Although designed as a performance safeguard, TLE errors are difficult to resolve: error messages provide no diagnostic insight, platform support is minimal, and existing debugging tools offer little help. As a result, many users abandon their submissions after repeated TLE failures.

This paper presents the first large-scale empirical study of TLE errors in online programming. We manually analyzed 1000 Codeforces submissions with TLE errors, classified their root causes, and traced how users attempted to fix them. Our analysis shows that TLE errors often arise not only from inefficient algorithms but also from infinite loops, improper data structure use, and inefficient I/O, challenging the conventional view that TLEs are purely performance issues.

Guided by these findings, we introduce Nettle, the first automated repair tool specifically designed for TLE errors, and Nettle-Eval, the first framework for evaluating TLE repairs. Integrating LLMs with targeted automated feedback generated by the compiler and test cases,  Nettle produces small, correct code edits that eliminate TLEs while preserving functionality. Evaluated on the same 1000 real-world cases, Nettle achieves a 98.5\% fix rate, far exceeding the strongest LLM baseline, and all of its repairs pass both Nettle-Eval and the platform’s official checker, confirming the reliability of our framework.

%% file: sec/intro.tex
\section{Introduction}
\label{sec:intro}

Online programming platforms such as Codeforces, LeetCode, and AtCoder play a crucial role in software development education and industry recruitment~\cite{springer2023,icpc2023}. These systems engage millions of users and evaluate code not only for functional correctness but also for computational efficiency. Each submission receives an automated verdict indicating the type of failure or success. Compared to the \texttt{Wrong Answer (WA)} verdict, where there is a bug in the program that causes incorrect outputs, the \texttt{Time Limit Exceeded (TLE)} verdict occurs when a submission exceeds the time budget (typically 1–2 seconds) despite producing no runtime or semantic errors. Table~\ref{tab:error_summary} shows that this error is particularly prevalent, appearing in 18\% of the 2.7~million submissions we sampled from Codeforces, and is especially frustrating for platform users.

Despite its frequency and relevance to learning practical efficiency constraints, TLE remains underexamined in programming education research. 
Unlike \texttt{Wrong Answer (WA)} or \texttt{Runtime Error (RE)}, which often provide outputs or stack traces, TLE offers only a one-line message, giving little guidance for debugging. Consequently, users are far less successful at resolving TLEs compared to other error types: while 31.7\% of WA submissions are eventually corrected, only 18.9\% of TLEs are fixed from our sampled 2.7 million submissions.
Unlike WA or RE, there is no partial output or error localization. Users are left to speculate whether the error comes from asymptotic inefficiency, data structure misuse, an infinite loop, I/O bottlenecks, or some other cause. Worse still, the test cases are hidden and unreproducible, making root cause analysis nearly impossible. TLE failures affect both novices and experts, often leading to excessive iteration. The median number of resubmissions from the first TLE to eventual success exceeds seven from our sampled Codeforces submissions, translating into wasted user time, reduced motivation, and blocked learning outcomes.

\begin{wraptable}{r}{0.55\linewidth}
\centering
\caption{Error type distribution in 2.7M Codeforces submissions. 
WA = Wrong Answer, TLE = Time Limit Exceeded, RE = Runtime Error, 
MLE = Memory Limit Exceeded, CE = Compilation Error. }
\label{tab:error_summary}
\begin{tabular}{lccccc}
\toprule
WA & TLE & RE & MLE & CE \\
1.91M & 432K & 163K & 67K & 196K \\
\bottomrule
\end{tabular}

\vspace{0.8em}

\begin{tikzpicture}
\pie[
   text=legend,                
   sum=auto,
   radius=1.6,
   hide number, 
   color={red!60, blue!50, green!50, orange!60, purple!50},
   before number=,             
   after number=,              
   every label/.style={text opacity=0}  
]{
  1910/WA,
  432/TLE,
  163/RE,
   67/MLE,
  196/CE
}
\end{tikzpicture}

\end{wraptable}

Existing tools fail to address TLE errors. Profilers and static analyzers lack access to platform-specific inputs or runtime behavior\cite{Resurrector}. Existing automated program repair (APR) techniques\cite{clara,Refazer} assume incorrect functionality and rely on failing test oracles, neither of which may be available for TLE. Similarly, recent advances in LLM-based code repair (e.g., CodeBERT\cite{codebert}, Codex\cite{Codex}) provide generic suggestions that ignore performance characteristics. As a result, programmers often resort to guesswork or brute-force rewriting or simply abandon the problem. Community forums offer little help, often responding with generic advice to optimize the algorithm or even misleading advice that ``a rule of thumb is that TLE errors are caused by inefficient algorithms.''
\footnote{\url{https://stackoverflow.com/questions/59227574/how-to-avoid-tle-when-array-is-used-to-store-1-105-integer-entries}}

From a research perspective, TLE represents a blind spot in existing tools and techniques. Unlike functional faults, TLE arises from the interaction of algorithmic complexity, data-dependent execution paths, language-level I/O costs, and subtle implementation inefficiencies, which together complicate diagnosis and repair. Prior work has pursued efficiency improvements, such as program transformations to optimize algorithms~\cite{anupam2025llmprogramoptimizationretrieval} and container substitutions guided by complexity analysis~\cite{Chengpeng22OOPSLA}, but these efforts evaluate isolated optimizations rather than offering a systematic, empirical account of TLE as a distinct failure mode. This gap motivates our study.

We present the first large-scale empirical study of TLE errors in online programming platforms. We manually analyze 1,000 real-world TLE submissions from Codeforces, a popular online programming platform. We manually categorize their root causes, and we identify five recurring patterns, including inefficient algorithms, infinite loops, and improper I/O. These findings provide the first taxonomy of TLE failures at scale and expose key shortcomings for addressing these errors using current programming and tooling practices.

Building on these insights, we design and implement \app, the \textbf{N}ew \textbf{E}xaminer \textbf{T}argeting \textbf{TLE} errors. Unlike prior systems that pursue efficiency gains only in semantically equivalent programs, \app directly addresses TLE cases that manifest as timeouts and may also contain functional errors. Our approach leverages large language models to repair code iteratively, generate candidate fixes, test them, and refine subsequent attempts based on prior feedback. To achieve this, our tool consists of two tightly coupled stages, \emph{reasoning} and \emph{judging}, which interact in an iterative loop to converge on efficient and correct repairs.

To evaluate \app, we construct a benchmark of 1,000 real-world TLE submissions from Codeforces and implement \appeval, an evaluation framework designed to assess both repair correctness and time-efficiency under strict resource limits. On this benchmark, \app achieves a fix rate of \overallaccuracy, substantially surpassing competitive baselines not only in repair success but also repair quality (small patch size). For example, Nettle reaches a 95.8\% fix rate, far above the strongest LLM-based baseline at 88.3\%, which approximates the best repair performance a user could achieve using general-purpose LLMs without tool-specific guidance. For repair quality, \app's edits are the same size as the baselines or even smaller.

To validate the reliability of \appeval itself, we submitted all repaired programs back to Codeforces for re-verification against the platform’s official ground truth. We found that every repair accepted by \appeval was also accepted by Codeforces, confirming the correctness of our local evaluation. In addition, we found one case where \appeval identified a \texttt{Wrong Answer} that Codeforces had marked as \texttt{Accepted}; after our report, the Codeforces maintainers confirmed the issue. This result further demonstrates that \appeval can reveal corner cases overlooked by the platform’s tests.
By design, \appeval adopts a conservative stance: because runtime environments and hardware conditions inevitably differ, it enforces stricter timing thresholds and flags borderline cases as \texttt{TLE} rather than accepting them. Together, these results demonstrate both the feasibility of automated TLE repair and the robustness of our evaluation framework, highlighting its potential to advance efficiency-aware program repair for online programming education and technical recruitment.

We make three core contributions:

\begin{itemize}
    \item We present the first large-scale empirical study of Time Limit Exceeded (TLE) errors, analyzing 1,000 Codeforces submissions and identifying five recurring root causes along with their distinctive repair challenges.  
    \item We introduce \app, the first automated repair tool explicitly targeting TLE errors, powered by large language models, together with \appeval, the first evaluation framework designed to validate both correctness and efficiency of TLE fixes systematically.
    \item We evaluate \app on 1,000 real-world TLE submissions and show that it achieves a fix rate of \textbf{\overallaccuracy}, consistently outperforming strong LLM-based baselines both by being more accurate and by producing more compact repairs with faster convergence.

\end{itemize}

The remainder of the paper is structured as follows.
Section~\ref{sec:motiv} walks through multiple motivating examples of
real TLEs.
Section~\ref{sec:tle-understanding} provides a study of TLE patterns. 
Section~\ref{sec:method} describes our
approach in detail. 
Section~\ref{sec:eval} provides experimental results on our dataset of Codeforces programs with TLE errors. 
Section~\ref{sec:discussion} details further discussions and limitations.
We discuss related work in Section~\ref{sec:related}.
Finally, we conclude the paper in Section~\ref{sec:conclusion}.

%% file: sec/motiv.tex
\section{Motivation}
\label{sec:motiv}

\subsection{What Are TLE Errors?}
\label{subsec:how}

Time Limit Exceeded (TLE) errors are a distinctive failure mode in online programming problems. Unlike syntax or semantic errors, TLE indicates that a program exceeds the platform's runtime constraints. This distinction makes TLE particularly elusive to diagnose and fix.

\begin{wrapfigure}{r}{0.55\textwidth}
\vspace{-1em}
\begin{minipage}{\linewidth}
\begin{lstlisting}[language=Python]
# TLE-inducing version
for i in range(len(s) - 1):
    if s[i:].startswith('AB'):
        abIndexes.append(i)

# Efficient fix
for i in range(len(s) - 1):
    if s[i] == 'A' and s[i + 1] == 'B':
        abIndexes.append(i)
\end{lstlisting}
\caption{TLE-inducing vs. optimized string ``AB'' check}
\label{lst:ab}
\end{minipage}
\vspace{-1em}
\end{wrapfigure}

Figure~\ref{lst:ab} illustrates a real-world TLE example from Codeforces\footnote{\url{https://codeforces.com/contest/550/submission/78561281}}. The original solution uses Python's \texttt{startswith} function on a sliced string to detect the pattern ``AB''. Although functionally correct, repeated slicing incurs $O(n^2)$ overhead on large strings. A more efficient solution compares adjacent characters directly.

Despite its apparent simplicity, this fix is nontrivial to uncover. Traditional fault-localization techniques fail because both inefficient and efficient versions execute identical control paths. Profilers can reveal hotspots but offer no concrete repair strategy, and online platforms typically hide the TLE-triggering inputs. Most critically, resolving the issue requires language-specific expertise that many learners, especially novices, lack: string slicing in Python has a hidden $O(n^2)$ cost. These challenges illustrate why TLE errors persist as a uniquely difficult class of programming failures.

\subsection{What Is Behind the ``Time Limit Exceed'' Error Message?}
\label{subsec:what}

\para{TLE Does Not Preclude Semantic Incorrectness.}
TLE is itself a vague error message. A common belief is that it indicates a semantically correct program that merely exceeds the time limit. To test this assumption, we randomly sampled and analyzed 1000 real-world TLE submissions from Codeforces by executing them locally under relaxed constraints (extended runtime) and with compiler optimizations, observing their outputs across varying input sizes.  

\begin{wraptable}{r}{0.45\textwidth}
\vspace{-1em}
\caption{Outcome distribution for 1000 real-world TLE submissions from Codeforces by executing them locally under relaxed constraints.}
\label{tab:mini_errors}
\centering
\begin{tabular}{lc}
\toprule
\textbf{Outcome} & \textbf{Count} \\
\midrule
Correct     & 152  \\
Wrong Answer     & 176 \\
TLE              & 516 \\
Other Verdicts     & 156 \\
\bottomrule
\end{tabular}
\vspace{-1em}
\end{wraptable}

As shown in Table~\ref{tab:mini_errors}, 516 programs produced correct outputs on small inputs but had TLE on larger inputs even under relaxed constraints. However, 176 programs still produced \emph{incorrect} outputs even when given additional runtime, revealing underlying semantic bugs. In addition, 152 programs passed only after enabling advanced compiler optimizations in \texttt{pypy3} , demonstrating that while such techniques offer modest performance gains, they fall far short of addressing the TLE problem in general.

\subsection{Why Are TLE Errors Challenging to Fix?}

\para{Limitations of Local Stress Testing.}
Stress testing generates large random inputs locally to evaluate runtime performance. While useful in certain settings, this approach is often insufficient for diagnosing TLE errors, as many inefficient algorithms behave acceptably on random inputs but degrade on edge cases (e.g., quicksort on sorted arrays). Furthermore, generating valid large-scale inputs for complex problems, such as connected graphs with tens of thousands of edges, demands problem-specific knowledge and tooling.

\para{Tooling Support Is Minimal.}  
Traditional debugging and program repair tools are ill-suited for TLE errors. General-purpose debuggers (gdb), refactoring assistants (e.g., Refactory\cite{Refactory}), and APR tools (e.g., GenProg\cite{GenProg}, Prophet\cite{prophet}, PyDex\cite{pydex}) rely on observable incorrect outputs. For example, APR tools typically locate faults by comparing passing and failing test cases and performing semantic patching. This approach fails for TLE. Programs may produce the correct output but simply run too slowly, making them indistinguishable from correct programs under conventional fault localization. 

In practice, users turn to forums like StackOverflow for help but find little actionable feedback:

\vspace{0.5em}
\begin{itemize}
    \item ``My Java code is getting Time Limit Exceeded, whereas the same logic in C++ doesn't.''\href{https://stackoverflow.com/questions/42785993}{[link]}
    \item ``Optimize Python code to read various inputs: Time Limit Exceeded.'' \href{https://stackoverflow.com/questions/67863740}{[link]}
    \item ``How to solve timeout termination issue (Time Limit Exceeded) while using for loop in JavaScript?'' \href{https://stackoverflow.com/questions/67863740/optimize-python-code-to-read-various-inputs-as-time-limit-exceeded}{[link]}
    
\end{itemize}
\vspace{0.5em}

Such questions often go unanswered or receive generic replies (``find an algorithm that doesn't need to do 100 million things''\footnote{\url{https://stackoverflow.com/questions/54496861/how-do-i-get-rid-of-the-tle-error-in-python}}) without diagnosing the actual bottleneck. Without tooling that connects performance traces to code locations, users are left guessing.

For example, in this Stack Overflow post, a user provided a detailed problem description and asked:\footnote{\url{https://stackoverflow.com/questions/73529959/how-can-i-overcome-this-tle}}

\begin{quote}
``How can I overcome a TLE error?''
\end{quote}

The only response was a code dump without any explanation. When the original poster followed up with:

\begin{quote}
``I'm still getting TLE, but otherwise I would actually prefer to try and fix my original code and understand how to reduce the computational time.''
\end{quote}

Nonetheless, no further reply was given. This case illustrates that TLE issues are often mishandled: an incorrect answer with no reasoning is worse than no answer at all, as it misleads learners and offers little educational value. Such examples highlight the urgent need for both an automated repair tool specifically designed for TLE errors and a trustworthy TLE validation framework.

%% file: sec/survey.tex
\section{Understanding TLE Errors}
\label{sec:tle-understanding}

To understand the underlying causes of TLE errors, we manually analyzed real-world submissions from the Codeforces platform. We selected representative problems across difficulty levels and examined failing programs that triggered TLE or related verdicts, together with their successful repairs. Our analysis uncovered five recurring failure patterns, each reflecting a distinct inefficiency mechanism.

\para{Inefficient Algorithm.}
The most frequent cause of TLE is excessive time complexity from brute-force enumeration.\footnote{\url{https://codeforces.com/contest/1/submission/24609270}} Example: 
\begin{lstlisting}[language=Python]
for n in range(2, 100000000, 2):
    if x * n >= a1:
        print(n)
        break
\end{lstlisting}
This linear scan performs billions of iterations.
The correct
approach would use mathematical rounding to compute the minimal multiplier in $O(1)$.

\para{Inefficient Handling.}
Redundant calls to expensive operations (e.g., strlen, memset) or repeatedly recompute or accumulate quantities that
could be derived directly.\footnote{\url{https://codeforces.com/contest/1/submission/24330765}} Example: 
\begin{lstlisting}[language=Python]
while plate_square < squares_square:
    plate_square = plate_square + (int(plate_size) ** 2)
    num_plates += 1
\end{lstlisting}
The code simulates tile accumulation by iterative addition rather than using direct division and
ceiling operations, incurring unnecessary runtime overhead.

\para{Improper I/O.} Format mismatches or type errors that result in incorrect loop bounds or
hanging input calls.\footnote{\url{https://codeforces.com/contest/1/submission/23397241}} Example: 
\begin{lstlisting}[language=Python]
n, m, a = input().split()
print((n//a + (n%a != 0)) * (m//a + (m%a != 0)))
\end{lstlisting}
Since the input tokens remain strings, integer operations such as \texttt{//} and \texttt{\%} fail or hang. Proper type conversion via \texttt{map(int, ...)} is essential before computation.

\para{Infinite Loop.}
Logical flaws in loop conditions often lead to nontermination.\footnote{\url{https://codeforces.com/contest/1/submission/22168665}} Example: 
\begin{lstlisting}[language=Python]
while mayor!=0 or mayor<0:
    mayor = mayor - c[2]
\end{lstlisting}
The use of logical \texttt{or} ensures the condition remains true even when \texttt{mayor} becomes negative. Combined with a zero or negative decrement, the loop never exits.

\para{Memory Access.}
Out-of-bounds writes or reads may produce undefined behavior,
sometimes manifesting as silent TLEs.\footnote{\url{https://codeforces.com/contest/490/submission/179836963}} Example: 
\begin{lstlisting}[language=C++]
int a[100];
for (int i = 0; i < n; i++) {
    cin >> a[i];  // overflow if n > 100
}
\end{lstlisting}
When \texttt{n} exceeds array bounds, memory corruption interferes with I/O buffers or loop control, producing a silent hang rather than a segmentation fault.

\para{Summary.}
TLE errors are multifaceted: they may stem from asymptotic inefficiency, unoptimized control flow, input mismanagement, or low-level memory misuse. These findings highlight that TLE is not purely an algorithmic failure but a spectrum of performance and correctness issues. Recognizing these fine-grained causes enables automated feedback systems to generate targeted, pedagogically meaningful guidance beyond simple runtime profiling.

%% file: sec/systemdesign.tex
\section{\app's System Design}
\label{sec:method}

\begin{figure}[htbp]
    \centering
    \includegraphics[width=0.8\linewidth]{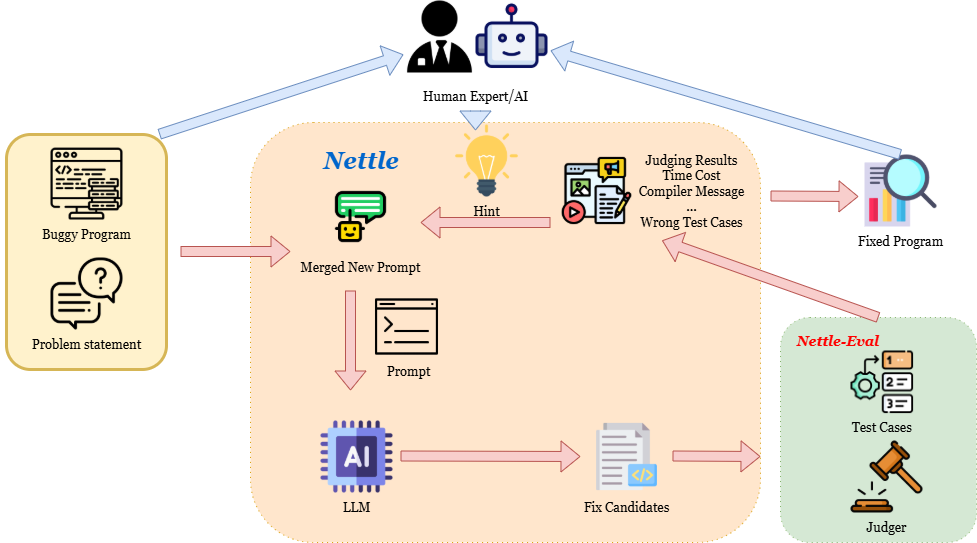}
    \caption{Overview of the system pipeline. The reasoning stage (Nettle) constructs structured prompts with optional hints and generates fix candidates via the LLM, while the judging stage (Nettle-Eval) executes them under resource limits and provides feedback (e.g., TLE, WA, test case) to guide further reasoning. }
    \label{fig:framework}
\end{figure}

As shown in Figure~\ref{fig:framework}, we propose an automated repair framework to address \emph{time-limit exceeded} (TLE) errors in programming assignments. The system takes as input an incorrect program submission $S$ for a given problem $P$, a set of visible input–output test cases, and a judger (Oracle). In the classical setting, a TLE program $S$ may produce an output on certain sample inputs (not necessarily correct), but fails to terminate within the time limit on other inputs. Our framework aims to transform $S$ into a corrected program $S'$, such that it (1) terminates within the time bounds enforced by the evaluator (provided by the problem description), (2) produces the expected outputs on all visible sample tests, and (3) generalizes to unseen inputs by both avoiding TLE and matching the expected outputs. To achieve this, our tool consists of two tightly coupled stages—\emph{reasoning} and \emph{judging}—which interact iteratively.

In the reasoning stage, an LLM-driven engine proposes \texttt{k} candidate repairs using prompts that integrate multiple sources of context. In the judging stage, a customized sandbox executes each candidate under strict resource limits to assess correctness and performance. The evaluator then produces structured feedback (e.g., which tests passed or failed, execution time, memory usage, error type) and returns it to the reasoning stage to guide the next iteration. Among the candidates, the one passing the most tests is retained while the others are discarded, effectively implementing a beam search with width one. The process repeats until a candidate passes all visible tests within the limits or the maximum iteration budget is reached. Inspired by how human programmers iteratively debug performance issues, the system alternates between reasoning, modification, and validation, with early termination ensuring efficiency when a valid patch is found before reaching the iteration limit.

\subsection{Reasoning Stage}

The reasoning module generates candidate repairs for TLE submissions. It aggregates context into a structured prompt for the LLM, including (1) the problem description $Q$, (2) the buggy code $S$, (3) diagnostic information (e.g., static analysis warnings, compiler messages, runtime profiles), (4) optional hints, and (5) prior repair history. The prompt formatter assembles these into a coherent query. For example, a prompt may include the problem description, the student’s code (with comments removed), and an instruction such as:  
\texttt{Task: Fix the TLE errors in the code so that it finishes within the time limit.}

Two types of hints are supported. A \emph{general} hint corresponds to an error class; for instance, for a TLE error: ``The code may contain an infinite loop, an inefficient algorithm, or slow input/output.'' General hints are produced automatically by \app, derived from the model's intermediate reasoning steps. A \emph{problem-specific} hint, in contrast, targets a particular assignment. For example, in Codeforces 102B\footnote{\url{https://codeforces.com/problemset/problem/102/B}}, a suitable hint is ``Use string input instead of integer input in Python.'' To ensure consistency and pedagogical value, we manually author these problem-specific hints and reuse them across all submissions for the same problem.

Once the prompt is prepared, the LLM (candidate synthesizer) generates \texttt{k} candidates per iteration ($C_t = \{P_{t,1}, \dots, P_{t,\texttt{k}}\}$), sampled with different parameters (temperature, top-$k$). This allows exploration of different optimization strategies: one candidate may reuse the precomputed values, another may use a more efficient data structure. The LLM is encouraged to make minimal edits (via instructions like ``fix the bug'' rather than ``rewrite the code''), preserving functionality. Each response is logged together with prompts and feedback, forming a memory across iterations. Feedback from the judging stage is merged into subsequent prompts, guiding the model away from repeated mistakes and towards convergence.  

\subsection{Judging Stage}
One key challenge in addressing TLE is ensuring that a repaired program can pass the rigorous correctness and performance checks imposed by online platforms such as Codeforces. To tackle this, we introduce the first evaluation framework, \appeval, centered on TLE resolution. \appeval serves as a correctness-and-performance oracle: it executes candidate patches against a set of representative inputs and uses verdicts such as \texttt{Accepted}, \texttt{Wrong Answer}, or \texttt{Time Limit Exceeded} to guide further repair in the reasoning stage.

The judging stage is implemented as a Docker sandbox running under strict resource limits. Candidates are compiled (or interpreted) and executed on the test suite with CPU time limit $T$ and memory limit $M$. Watchdogs enforce timeouts: if execution exceeds $T \cdot f$ seconds (with $f$ a factor, typically 3--5), the process is killed and CPU time recorded. Memory is constrained to $M$ MB, with overuse labeled Memory Limit Exceeded. The sandbox is deterministic, side-effect free, and reproducible. Moreover, this isolation prevents possible harmful code generated by LLMs from affecting the host system.  

The judger produces a structured verdict with standard contest labels: AC (Accepted), WA (Wrong Answer), RE (Runtime Error), CE (Compilation Error), TLE (Time Limit Exceeded), and MLE (Memory Limit Exceeded). Diagnostics include failing test indices (e.g., ``test 7 timed out''), execution time, memory usage, and captured error traces. Functional correctness is assessed by exact output comparison. \appeval supports an optional feature called \emph{special judges} that can check for more complex properties, such as inclusion in multivalued sets.  For example, Codeforces~584A \footnote{\url{https://codeforces.com/problemset/problem/584/A}} requires any $n$-digit number divisible by $t$. We implement a custom checker as follows to support multiple valid outputs: 

\begin{lstlisting}[language=Python, basicstyle=\ttfamily\small, caption={Special judge for Codeforces 584A.}]
def _584A(file1, file2, file3):
    def trans(s):
        s = list(map(lambda s: s.rstrip(), s.split('\n')))
        if s[-1] == '':
            s = s[:-1]
        return s

    with open(file1) as f:
        f1 = trans(f.read())
    with open(file2) as f:
        f2 = trans(f.read())
    if not len(f1) == 1:
        return True
    f1 = f1[0]
    f2 = f2[0]
    if int(f2) == -1:
        return f1 != f2

    with open(file3) as f:
        line = f.readline()
        n, t = map(int, line.split())

    if not f1.isdigit():
        return True
    if len(f1) != n:
        return True
    if f1[0] == '0':
        return True
    s1 = int(f1)
    if s1 % t == 0:
        return False
    else:
        return True
\end{lstlisting}

After passing the judging stage, the output is collected, analyzed, and encoded into natural-language prompts for the next reasoning round. For example:

\noindent\textbf{After TLE (on visible cases):}
\begin{lstlisting}[basicstyle=\ttfamily\small]
Time Limit Exceeded. The program may contain 
an infinite loop or an inefficient algorithm.
\end{lstlisting}

\noindent\textbf{After WA (on visible cases):}
\begin{lstlisting}[basicstyle=\ttfamily\small]
Wrong Answer. For input X, expected Y but got Z. 
Check the logic and ensure correctness.
\end{lstlisting}

\noindent\textbf{After RE (on visible cases):}
\begin{lstlisting}[basicstyle=\ttfamily\small]
Runtime Error. The program crashed (e.g., null pointer). 
Fix the underlying runtime issue.
\end{lstlisting}

Therefore, the system functions as an iterative loop: the reasoning stage proposes candidate patches, the judging stage executes and evaluates them, and the candidate passing the largest number of tests is retained for the next round (ties are broken randomly). Feedback from the evaluator then guides the next iteration. The process continues until a patch $S'$ passes all tests within the resource limits or the maximum depth budget is exhausted. Empirically, successful repairs are typically found within only a few iterations. By systematically combining LLM-based synthesis with rigorous sandbox evaluation, our framework ensures that only patches that \emph{truly resolve TLE while preserving correctness} are accepted, closely mirroring how human programmers iteratively debug performance issues.

To sum up, this judger design ensures that a repaired program passes a rigorous checking that emulates that of an online platform. We have wrapped and released our judger as a standalone gym, providing a dedicated testing framework to support future research on TLE.

%% file: sec/evaluation.tex
\section{Evaluation}
\label{sec:eval}

We evaluate \app by answering the following research questions:

\begin{itemize}
  \item \textbf{RQ1: Repair Effectiveness.} 
  How effective is \app in repairing real-world TLE errors compared to state-of-the-art LLM-based repair baselines, in terms of success rate and repair size?
  
  \item \textbf{RQ2: Role of Contextual Information.} 
  How does the availability of different contextual signals (e.g., general error reasons, problem-specific hints) influence the repair performance of \app?

  \item \textbf{RQ3: Effectiveness of \appeval.} How effective is \appeval in assessing repair candidates within \app? In particular, is the proposed change a genuine repair of the original code, or merely a plausible but incorrect repair?

\end{itemize}

\subsection{Implementation Setup}
We implemented \app in Python, leveraging open-source libraries for code analysis and repair. The system is designed to address both syntactic and semantic TLE errors in a unified framework.  

For candidate ranking, we used log-probability scoring from the language model component (WizardCoder, fine-tuned from OpenAI’s CodeLlama). In each iteration, the model generates 10 candidates, which are then evaluated against the visible test suite. Among these, the candidate passing the largest number of tests is retained while others are discarded, ensuring that only the strongest repair is carried forward to the next iteration. This iterative process continues until either a candidate passes all visible tests or the maximum iteration budget is exhausted.  

We experimented with Qwen2.5-Coder-32B as our main engine, and compared it against Llama-3 (33B Instruct) and WizardCoder, representing a state-of-the-art LLM baselines.

Experiments were run on Ubuntu 22.04, Intel i9 CPU, 64 GB RAM, and dual Nvidia P40 GPUs.

\subsection{Benchmarks}
We constructed a new benchmark of 1,000 real-world TLE submissions drawn from Codeforces. We selected 10 problems (difficulty 800–1000) with the highest number of submissions and no image-based descriptions. For each problem:
\begin{enumerate}
    \item We randomly sampled 50 submissions that were eventually repaired by the user (with a ground truth).
    \item We randomly sampled 50 submissions that users eventually abandoned (challenging cases).
\end{enumerate}

Notably, many of these submissions exhibit not only efficiency bottlenecks but also additional semantic issues (e.g., hidden wrong-answer or runtime errors that appear once performance is improved). This makes the benchmark more realistic and challenging, and it allows us to demonstrate that \app can repair both \emph{pure} TLE cases and \emph{mixed} TLE+functional errors.  

This yields a balanced dataset of 500 \emph{eventually repaired} and 500 \emph{never repaired} programs.  
All benchmarks, code, and Codeforces submission history will be made available during the artifact evaluation phase, to be unblinded upon paper acceptance.
To show the dataset statistics, in Table~\ref{tab:tle_errors},
we manually categorized each submission into the root-cause categories described in Section~\ref{sec:tle-understanding}.

\begin{table}[t!]
\centering
\caption{Distribution of TLE root causes in our dataset.}
\label{tab:tle_errors}
\begin{tabular}{@{}ll@{}}
\toprule
Error Type                          & Count \\ \midrule
Inefficient algorithm               & 473 \\
Inefficient Handling & 218 \\
Improper I/O         & 164 \\
Runtime errors (e.g., Endless Loop, Memory Access) & 84 \\
Completely wrong / irrelevant code  & 13 \\
Other (e.g., input format stalls)   & 132 \\ \bottomrule
\end{tabular}
\end{table}

\subsection{RQ1: Comparison with Baselines}
As a baseline, we simulate the repair process that a typical user would perform by directly prompting a large language model (LLM) to fix the code, without using any additional repair tools. This baseline isolates the model’s intrinsic ability to generate correct repairs.

\para{Evaluation Metrics.}
We adopt an evaluation protocol that closely emulates the Codeforces judging process. The test suite for each problem is partitioned into two subsets: an initial 10\% used during repair and a held-out 90\% used for final validation. During the repair process, candidates are only executed against the 10\% subset. If a candidate fails on any of these tests, it is immediately discarded. Once the repair loop produces a candidate that passes all visible tests, this candidate is then evaluated on the remaining 90\%.  

We consider a repair successful if the final candidate passes all held-out cases with the correct expected outputs and without TLE errors. This staged setup allows early pruning of invalid repairs while still providing rigorous validation against unseen inputs. In our experiments, we verified that 100\% of the fixes flagged as correct by this protocol were indeed correct.  

\para{Results.}
Across the benchmark, \app consistently surpasses all baselines, regardless of the underlying LLM (Table~\ref{tab:rq1_results}).
With Qwen2.5-Coder, \app attains a 98.5\% repair success rate with a small average edit distance, demonstrating both accuracy and effectiveness.
To ensure fairness, all systems followed the same repair budget: ten candidates were generated per iteration, the best candidate was retained at each step, and the process was repeated for up to five iterations.

\begin{table}[t]
\centering

%

\caption{Repair success rates across system variants (10 problems × 100 submissions each).}
\label{tab:rq1_results}
\begin{tabular}{@{}lccc@{}}
\toprule
System & Repair Rate  & Average Edit Distance \\ \midrule
baseline (wizardcoder) & 44.0\% & 59.1  \\
baseline (Llama-3) & 73.0\% & 61.2  \\
baseline (Qwen2.5-Coder) & 88.3\% & 58.5  \\
\app (WizardCoder) & 95.7\% & 44.8 \\

\app (Llama-3) & 94.5\% & 48.9 \\

\textbf{\app (Qwen2.5-Coder)} & \textbf{98.5\%} & 52.6  \\\bottomrule
\end{tabular}
\end{table}

\subsection{RQ2: Role of Contextual Information}
We next investigate how providing contextual information affects repair outcomes. We consider two forms of hints:
\begin{itemize}
    \item \textbf{General hint} (e.g., “This code likely times out due to inefficient input handling”).
    \item \textbf{Problem-specific hint} (e.g., “For Problem 102B, convert inputs to strings to avoid repeated parsing overheads”).
\end{itemize}

\begin{table}[h]
\centering
\caption{Impact of hints on repair rates by hint type and model .}
\label{tab:rq2_results_by_hint}
\resizebox{\linewidth}{!}{
\begin{tabular}{@{}l|ccc|ccc|ccc@{}}
\toprule
\multirow{2}{*}{Problem} 
& \multicolumn{3}{c|}{No Hints} 
& \multicolumn{3}{c|}{General hint} 
& \multicolumn{3}{c}{Problem-Specific Hint} \\ 
\cmidrule(lr){2-4}\cmidrule(lr){5-7}\cmidrule(lr){8-10}
 & Llama-3 & Wizardcoder & Qwen2.5-Coder 
 & Llama-3 & Wizardcoder & Qwen2.5-Coder 
 & Llama-3 & Wizardcoder & Qwen2.5-Coder \\ 
\midrule
339B & 44\% & 8\% & 80\% & 83\% & 99\% & 100\% & 84\% & 99\% &  100\%\\
379A & 76\% & 28\% & 75\% & 97\% & 94\% & 92\% & 98\% & 94\% & 96\% \\
486A & 7\% & 6\% & 90\% & 76\% & 94\% & 98\% & 87\% & 95\% & 100\% \\
490A & 79\% & 23\% & 83\% & 97\% & 97\% & 98\% & 97\% & 97\% & 99\% \\
584A & 78\% & 42\% & 92\% & 96\% & 100\% & 100\% & 97\% & 100\% & 100\% \\
617A & 93\% & 74\% & 98\% &  100\% & 100\% & 100\% & 100\% & 100\% & 100\% \\
791A & 94\% & 78\% & 99\% & 99\% & 99\% & 100\% & 100\% & 100\% & 100\% \\
977A & 97\% & 85\% & 95\% & 97\% & 99\% & 99\% & 98\% & 99\% & 99\% \\
1A   & 99\% & 76\% & 100\% & 100\% & 100\% & 100\% & 100\% & 100\% & 100\% \\
102B & 63\% & 20\% & 71\% & 100\% & 75\% & 98\% & 100\% & 96\% & 97\% \\
\bottomrule
\end{tabular}}
\end{table}

Table~\ref{tab:rq2_results_by_hint} shows that hints significantly improve repair rates, especially for harder problems (e.g., 102B \footnote{\url{https://codeforces.com/problemset/problem/102/B}}). 
Lightweight contextual cues substantially boost repair success, in some cases more than doubling accuracy on difficult problems (e.g., 102B). Problem-specific hints yield the largest improvements but need to be designed only once per problem, making them a low-overhead addition. We also observe that such hints benefit general-purpose models like LLama more than code-specialized models like WizardCoder, highlighting that contextual signals can compensate for weaker domain specialization.

Finally, we isolate the effect of the number of candidates per iteration (fan-out, $C$) and the total number of repair iterations ($D$). Table~\ref{tab:ablation_comparison} shows that increasing either factor alone yields modest gains, but combining both yields the largest improvements. Breadth (candidate diversity) and depth (iterative refinement) are complementary; the best performance emerges when both are maximized.

\begin{table}[h]
\centering
\caption{Effect of candidate pool size ($C$) and repair depth ($D$) on repair rates across different code repair models. $C$ denotes the number of candidate fixes generated per round (i.e., pool size), while $D$ denotes the number of iterative repair rounds (i.e., repair depth). The analysis focuses on challenging repair instances to highlight the benefit of increasing $C$ and $D$. Bold values indicate superior performance.}
\label{tab:ablation_comparison}
\begin{tabular}{@{}l|ccc@{}}
\toprule
\textbf{Configuration} & \textbf{Wizardcoder} & \textbf{LLama-3} & \textbf{Qwen2.5-Coder}   \\ 
\midrule
$C{=}1, D{=}1$ & 44.0\%  & 73.0\% & 88.3\%  \\
$C{=}1, D{=}5$ & 58.0\% &  87.9\% & 89.0\%  \\
$C{=}5, D{=}1$ & 76.0\% & 78.4\% & 90.5\% \\
$C{=}5, D{=}5$ & 95.7\% &94.5\% & \textbf{98.5\%}  \\
\bottomrule
\end{tabular}
\vspace{0.3cm}

\begin{minipage}{\textwidth}
\footnotesize
\end{minipage}
\end{table}

\para{Why Existing Semantics Repair Technique Cannot Fix TLE.}
Time Limit Exceeded (TLE) errors fundamentally differ from the bugs that semantics-based repair systems such as Refactory\cite{Refactory} and Clara\cite{clara} are designed to handle. These tools target short, single-function programs and rely on local control-flow mutations while ignoring input/output, but TLE errors almost always arise from \emph{global inefficiency} in I/O handling or algorithmic design. In our dataset, 20--30\% of TLE cases come directly from input bottlenecks---for example, in Codeforces 102B\footnote{\url{https://codeforces.com/problemset/problem/102/B}}, 
\texttt{n = int(input())} 
times out on very large numbers, and the correct fix requires string-based parsing, which these tools cannot even access because they do not repair I/O code. More critically, TLE fixes often require algorithmic redesign that produces code structurally unrelated to the buggy version: in Codeforces 490A\footnote{\url{https://codeforces.com/problemset/problem/490/A}}, the buggy greedy-loop solution and the correct array-based grouping solution share almost no control-flow similarity, making transplantation impossible. Finally, because Refactory judges only output equivalence, it may wrongly accept programs that succeed on small cases but still gets TLE on large inputs, and slow TLE executions shrink the search budget from thousands of iterations to just one. In short, TLE repair requires reasoning about \emph{time complexity} and \emph{input performance}, which semantics-based repair is incapable of addressing.

\subsection{RQ3: Effectiveness of \appeval}
\label{sec:rq3}
To assess the reliability of our evaluation framework, we compared the outcomes of \appeval (Qwen2.5-Coder-32B) with the official Codeforces verdicts on a sample of 1000 submissions. Remarkably, the two verdicts matched in all but 15 cases. Even more encouragingly, these mismatches consistently reflect that \appeval is \emph{stricter} than Codeforces: the 10 programs flagged by our framework as failing were all accepted by Codeforces, yet we found no cases where Codeforces rejected a program that our framework accepted.
Closer inspection reveals the following breakdown of the 10 mismatches. Of these, one was judged \texttt{Wrong Answer (WA)} by our local evaluator, and nine were judged \texttt{TLE}. All 10 of these submissions were nevertheless labeled \texttt{Accepted (AC)} by Codeforces. This asymmetry suggests that our framework exposes corner cases overlooked by the platform’s public test data. In particular, the extra \texttt{TLE} verdicts can be attributed to natural variance in runtime measurement across environments, while the additional \texttt{WA} cases imply that our test inputs are stronger or more diverse than those used in Codeforces, uncovering errors that would otherwise go unnoticed. Notably, the \texttt{WA} case was successfully reported to Codeforces, further validating the accuracy and value of \appeval.
A concrete example illustrates this case. Consider the following repaired candidate for a digit-summation task \footnote{\url{https://codeforces.com/problemset/problem/102/B}}:
\begin{lstlisting}[language=Python]
def sum_of_digits(s):
    return sum(int(digit) for digit in s)
 
def count_operations_to_single_digit(n_str):
    count = 0
    while len(n_str) > 1:
        n_str = str(sum_of_digits(n_str))
        count += 1
    return count
 
# Read the input as a string
n_str = input().strip()

if  0 <= int(n_str) <= 10**100000:
    result = count_operations_to_single_digit(n_str)
    print(result)
else:
    print("Input out of range")
\end{lstlisting}
While this code appears correct, the use of \texttt{int(n\_str)} to check numeric bounds introduces a hidden scalability bottleneck.
The call converts the entire string into an integer—an operation that becomes prohibitively expensive or even infeasible when the input length approaches the declared upper bound (up to $10^{100000}$ digits).
On Codeforces, where inputs are typically smaller, this issue remains undetected and the program is marked as \texttt{AC}.
Under \appeval, however, the larger and more realistic input distribution triggers a stall or timeout, leading to a \texttt{TLE}.
This example highlights how subtle implementation choices can interact with runtime conditions to produce divergent verdicts, and how \appeval uses its broader input coverage to successfully expose such latent inefficiency bugs.
Overall, these results demonstrate that \appeval is highly effective: it matches Codeforces in more than 97\% of cases, and in the few disagreements, our framework errs on the side of rigor. Far from being a weakness, this stricter stance highlights an important advantage: \appeval can surface hidden flaws and performance issues that would otherwise be misclassified as correct. This property makes \appeval particularly well-suited for educational settings, where learners benefit more from precise and conservative feedback than from overly permissive acceptance.

%% file: sec/discussion.tex
\section{Discussion}
\label{sec:discussion}
We conclude by discussing qualitative aspects of repair quality that quantitative metrics alone cannot capture.

\subsection{Repairs vs. Full Rewrites}

A key nuance in evaluating automated repair is distinguishing \emph{incremental fixes} from \emph{full rewrites}. Consider the following case in Figure~\ref{fig:rewrite-case}.

\begin{figure}[h!]
\centering
\begin{minipage}[t]{0.45\linewidth}
\begin{lstlisting}[language=Python]
n=int(input())
c=0
while n>9:
    a=n
    s=0
    while a!=0:
        m=a%10
        s+=m
        a=a//10
    n=s
    c+=1
print(c)
\end{lstlisting}
\end{minipage}
\hfill
\begin{minipage}[t]{0.45\linewidth}
\begin{lstlisting}[language=Python]
n=input()
c=0
while len(n)!=1:
    n=[int(i) for i in n]
    n=str(sum(n))
    c+=1
print(c)
\end{lstlisting}
\end{minipage}
\caption{The user's ``fix'' is a complete rewrite: the original solution  uses integer division and modulus with nested loops, while the new code abandons this strategy and re-implements the task using string processing.}
\label{fig:rewrite-case}
\end{figure}

The original program\footnote{\url{https://codeforces.com/contest/102/submission/40030614}} attempted digit extraction with integer division and modulus, but the nested loops led to inefficiency and ultimately a \texttt{TLE} verdict. The user's fix\footnote{\url{https://codeforces.com/contest/102/submission/40032105}}, however, did not refine this approach; it discarded it entirely by switching to string conversion and list comprehensions.  

Such rewrites are fundamentally distinct from true repairs. Our system is designed to preserve the original problem-solving intent and propose actionable corrections (e.g., reducing redundant computation, adjusting loop bounds). By contrast, the user’s new submission represents a complete replacement. It should not be interpreted as a limitation of \app, but rather as an instance of programmers abandoning one strategy and starting over with another, a common behavior in online programming.  

This distinction underscores the pedagogical value of automated repair: incremental edits scaffold learning and help programmers refine existing solutions, whereas wholesale rewrites obscure intent and provide little insight into the repairability of the original code.

\subsection{High-Quality Repair Example (Problem 1A, Submission 10700220)}

\begin{figure}[t]
\centering
\begin{minipage}[t]{0.45\linewidth}
\begin{lstlisting}[language=Python]
# Original buggy code (TLE)
mna=[int(i) for i in input().split(' ')]
m=mna[0]
n=mna[1]
a=mna[2]
auxm=0
auxn=0
kn=0
km=0
while auxm<m:
    auxm=auxm+a
    km=km+1
while auxn<n:
    auxn=auxn+a
    kn=kn+1
out=km*kn
print(out)
\end{lstlisting}
\end{minipage}
\hfill
\begin{minipage}[t]{0.45\linewidth}
\begin{lstlisting}[language=Python]
# Tool-generated fix (O(1))
mna=[int(i) for i in input().split(' ')]
m=mna[0]
n=mna[1]
a=mna[2]
km=m//a
kn=n//a
if m%a!=0:
    km=km+1
if n%a!=0:
    kn=kn+1
out=km*kn
print(out)
\end{lstlisting}
\end{minipage}
\caption{Our tool transforms two inefficient linear loops ($O(m/a+n/a)$) into a direct ceiling formula using integer division and conditional increment ($O(1)$).}
\label{fig:high-quality-fix}
\end{figure}

The example\footnote{\url{https://codeforces.com/contest/1/submission/10700220}} in Figure~\ref{fig:high-quality-fix} highlights the quality of our system’s repairs. The buggy submission relied on linear accumulation loops, leading to $O(m/a+n/a)$ complexity and repeated \texttt{TLE}. Our tool replaced this with integer division plus conditional rounding up, the mathematically standard ceiling formula. This change guarantees correctness, since it matches the official editorial solution and handles both divisible and non-divisible boundaries. It improves complexity by reducing the algorithm to constant time, fully avoiding timeout risks. It enhances readability, as the repaired code directly reflects the intended ``rows $\times$ columns'' geometry. Finally, it improves robustness by eliminating reliance on intermediate accumulations, thereby reducing the chance of implementation errors or overflow. Together, these properties demonstrate that our system generates fixes that are not only valid, but also optimal, elegant, and pedagogically aligned with canonical human solutions.

\subsection{Iterative Repair Example (Problem 339B, Submission 35093821) }

The example\footnote{\url{https://codeforces.com/contest/339/submission/35093821}}  in Figure~\ref{fig:repair-pipeline} illustrates how \app enables iterative refinement of repairs. Starting from a TLE submission, we produced an intermediate repair that improved efficiency but violated the semantics (WA), before converging on a correct solution (AC). This trajectory highlights the value of repair pipelines: debugging is rarely a one-shot process, and effective evaluation supports progressive improvement.  

\begin{figure*}[t]
\centering
\begin{subfigure}[t]{0.30\linewidth}
\begin{lstlisting}[language=Python]
n, m = list(map(int, input().split()))
task = list(map(int, input().split()))
count = 0
cur = 1  # Start from house 1
for i in task:
    while cur != i:
        if cur == n:  # wrap to 1
            cur = 1
        else:
            cur += 1
        count += 1
print(count)
\end{lstlisting}
\caption{Attempt 1: Original buggy code (TLE, O(nm))}
\end{subfigure}
\hfill
\begin{subfigure}[t]{0.30\linewidth}
\begin{lstlisting}[language=Python]
n, m = list(map(int, input().split()))
task = list(map(int, input().split()))
count = 0
cur = 1

for i in task:
    dist_clockwise = (i - cur) % n
    dist_c_clockwise = (cur - i) % n
    # Error: may move counterclockwise
    if dist_clockwise <= dist_c_clockwise:
        count += dist_clockwise
    else:
        count += dist_c_clockwise
    cur = i
print(count)
\end{lstlisting}
\caption{Attempt 2: WA (counterclockwise shortcut)}
\end{subfigure}
\hfill
\begin{subfigure}[t]{0.30\linewidth}
\begin{lstlisting}[language=Python]
n, m = list(map(int, input().split()))
task = list(map(int, input().split()))
count = 0
cur = 1  # initial position is house 1

for i in task:
    if cur <= i:
        dist = i - cur  # forward
    else:
        dist = n - cur + i  # wrap around
    count += dist
    cur = i

print(count)
\end{lstlisting}
\caption{Attempt 3: AC (clockwise-only distance)}
\end{subfigure}
\caption{Repair pipeline for Problem 339B: (a) TLE due to simulation, (b) WA from invalid counterclockwise moves, (c) AC by encoding the correct clockwise distance.}
\label{fig:repair-pipeline}
\end{figure*}

The progression from Attempt~1 $\rightarrow$ Attempt~2 $\rightarrow$ Attempt~3 shows how \appeval not only distinguishes genuine repairs from overfitting rewrites, but also supports an \emph{iterative design}. Rather than rejecting intermediate fixes outright, the framework helps programmers understand why a solution fails (efficiency vs.\ semantics) and guides them toward principled, optimal solutions.

%% file: sec/related.tex
\section{Related Work}
\label{sec:related}

\para{Time-Limit-Exceeded (TLE) Errors.}
The software engineering and programming languages communities have long sought to build tools that assist learners in diagnosing and correcting errors in online programming environments.
Prior work spans a wide range of tasks, including error localization~\cite{5477098,GlassmanLCM16,SHErrLoc}, automated repair~\cite{Autograder,wang2018dynamic,Clef}, and intelligent feedback generation~\cite{Autograder,clara,SAR,Refactory,FAPR,Cafe,Li22generating}.
However, the vast majority of these efforts target programs with observable \emph{wrong-answer} outcomes, where the output diverges from a reference solution under the same inputs.
In contrast, \emph{time-limit-exceeded (TLE)} errors~\cite{halim2013competitive} represent a fundamentally different failure mode: the program is computationally inefficient, and thus far more challenging for learners to diagnose or repair.
While a few studies have explored performance-related bug repair~\cite{Caramel,Cachetor}, they focus on narrowly defined patterns such as too many loops or repeated computations. These approaches cannot address the broader algorithmic inefficiencies and improper I/O handling that dominate real-world TLE cases.

\para{Automated Feedback Generation.}
Over the past decade, the problem of automatically generating feedback for incorrect programming assignments has attracted substantial attention from the software engineering and computing education communities~\cite{Zimmerman15ASE,QLOSE,Refazer,YiFSE17,Paassen_2018,wang2018dynamic,Cafe,Li22generating,yao2025traininglanguagemodelsgenerate}.
Early systems~\cite{Autograder,CoderAssist} required manual intervention from instructors to guide feedback creation.
More recent approaches~\cite{clara,SAR,Refactory,FAPR} achieve automation by leveraging large repositories of student submissions to generate feedback. 
Despite these advances, existing tools are ill-suited for handling programs that fail due to TLE errors. Some techniques~\cite{BIFI,FAPR} operate only on syntactic faults, whereas programs having TLE errors are syntactically correct. Others, including SemCluster and Refactory~\cite{SemCluster,Refactory}, focus on semantic errors and rely heavily on dynamic analysis. Dynamic instrumentation becomes ineffective when every test input results in a timeout, leaving no successful execution to observe.
More fundamentally, most feedback generators equate correctness with input-output equivalence and ignore runtime efficiency. Consequently, they cannot diagnose or repair programs that are functionally correct but computationally inefficient, the top class of problems targeted in this work.

\para{LLM-based Automated Program Repair.}
Automated program repair (APR) has been extensively studied over the past decade. Early research primarily explored \emph{heuristic-based} approaches~\cite{APRGoues13,APRGoues19,GenProg,RandomQi,PAR} and \emph{semantics-based} techniques~\cite{Angelix,MechtaevNNGR18,Verifix}, which rely on mutation heuristics or symbolic reasoning to synthesize patches.
More recently, \emph{learning-based} methods~\cite{prophet,genesis,SKP,BIFI,Hoppity,Santolucito22} have become prominent by learning repair patterns from historical patches, yielding higher fix quality and generalization.
With the advent of large language models, \emph{LLM-driven} repair systems~\cite{LaMirage,DeepMerge,gmerge,karimipour2025llmbasedrepairstaticnullability,Danfeng2025,zeng2025bridgingeditinggapllms,rondon2025evaluatingagentbasedprogramrepair,shao2025llmscorrectlyintegratedsoftware,si2025viscratchusinglargelanguage} have rapidly emerged, demonstrating strong capabilities in generating human-like patches across diverse languages and domains.
Despite these advances, existing APR techniques are suited for isolated bugs in large codebases. In contrast, TLE errors on online programming platforms often require algorithm-level redesign rather than localized edits. This distinction makes TLE repair a substantially different and underexplored challenge for LLM-based systems.

\para{Online Programming Assignments.}
Recent research on online programming platforms has focused on enhancing the overall learner experience through multiple directions.
First, natural-language–based feedback systems generate personalized explanations and hints for student code~\cite{phung2023generatinghighprecisionfeedbackprogramming}.
Second, automated test-case generation frameworks seek to construct comprehensive test suites, thereby enhancing the validity and reliability of automated grading~\cite{Cafe}.
Third, several studies explore how to design more effective and engaging programming assignments~\cite{LLMAgainst}.
Finally, recent empirical work investigates how automated feedback affects learning outcomes and user behavior~\cite{kiesler2022exploratoryanalysisfeedbacktypes}.
Our work is complementary to these efforts: it demonstrates that the challenging problem of repairing time-limit-exceeded (TLE) submissions on online programming platforms can be effectively addressed through \app.

%% file: sec/conclusions.tex
\section{Conclusion}
\label{sec:conclusion}

This work provides a systematic understanding of Time Limit Exceeded (TLE) errors in online programming platforms and demonstrates the feasibility of automated support for their diagnosis and repair. Through a large-scale empirical study of 1,000 Codeforces submissions, we identify key root causes, including not only algorithmic inefficiency but also semantic errors such as infinite loops, poor data structure choices, and inefficient I/O. These findings challenge prevailing assumptions about TLE and uncover concrete opportunities for performance-aware feedback.

Building on this analysis, we introduce \app and \appeval, which together enable automated repair of TLE errors at scale. Our evaluation shows that targeted, LLM-driven feedback can outperform generic repair models in both accuracy and repair quality. These results point to a promising direction for future tools that reduce trial-and-error, improve the programming experience, and foster deeper algorithmic insight—especially for learners navigating complex problem-solving tasks.